# Mean field theory and Monte Carlo simulation of Phase transitions and Magnetic Properties of a tridimensional Fe$_7$S$_8$ Compound

*S. Benyoussef$^{(1)}$, Y. EL Amraoui$^{(1,2)}$, H. Ez-Zahraouy$^{(1)}$, D. Mezzane$^{(3)}$, Z. Kutnjak$^{(4)}$, I. A. Luk'yanchuk$^{(5)}$, M. EL Marssi$^{(5)}$*

$^{(1)}$ LaMCScI, Faculté des Sciences, Université Mohamed V, Rabat, Morocco.
$^{(2)}$ Département des Matériaux et Génie des procédés, Ecole Nationale d'Arts et Métiers, Université Moulay Ismail, Meknès, Morocco.
$^{(3)}$ Laboratoire de la Matière Condensée et Nano structures LMCN, FSTG, Université CadyAyyad, Marrakech, Morocco.
$^{(4)}$Jožef Stefan Institute, Ljubljana IJS, Slovenia.
$^{(5)}$University of Picardy, Laboratory of Cond. Mat. Physics, Amiens, France.

**Abstract**

The structural, electronic and magnetic properties of Fe$_7$S$_8$ material have been studied within the framework of the ab-initio calculations, the mean field approximation (MFA) and Monte Carlo simulation (MCS). Our study shows that two forms of the iron atoms, Fe$^{2+}$ with spin S=2, and Fe$^{3+}$ with spin $\sigma$=5/2 are the most probable configurations. A mixed Ising model with ferromagnetic spin coupling between Fe$^{2+}$ and Fe$^{3+}$ ions and between Fe$^{3+}$ and Fe$^{3+}$ ions, and with antiferromagnetic spin coupling between Fe$^{2+}$ ions of adjacent layers has been used to study the magnetic properties of this compound. We demonstrated that the magnetic phase transition can be either of the first or of the second order, depending on the value of the exchange interaction and crystal field. The presence of vacancies in every second iron layer leads to incomplete cancellation of magnetic moments, hence to the emergence of the ferrimagnetism. Anomalies in the magnetization behavior have been found and compared with the experimental results.

**Keywords:** Pyrrhotite Fe$_7$S$_8$ · Ferromagnetic · Ferrimagnetic · Antiferromagnetic · Ising model · Mixed spin · Mean field approximation · Ab-initio calculation · Monte Carlo simulation.

## I. Introduction

Monoclinic Pyrrhotite compound Fe$_7$S$_8$ is the natural iron sulfide founded in the Earth's crust[1]. This Pyrrhotite contributes to rock magnetization and used in paleomagnetic studies as an indicator for the reversals magnetic field of the Earth [2]. In the laboratory, the Fe$_7$S$_8$ was synthesized by solid-state reactions in evacuated quartz tubes [1] and also by the silica tube technique [3].
The nanobiscuit structure of Fe$_7$S$_8$@C is composed of the core Fe$_7$S$_8$ material and the shell carbon layer, which exhibits the high specific capacities, excellent rate capabilities, and stable cycling. Previous studies clearly prove that these materials with well-designed biscuit structure can be used as anodes for lithium-ion and sodium-ion batteries [4,5]. In ref. [6] the Fe$_7$S$_8$ nanomaterial was produced and exploited for the cancellation of Pb$^{2+}$ and Cu$^{2+}$ ions from aqueous solutions, in which the rate of the percentage removal increases with increasing of the temperature. Pyrrhotite was also used to remove arsenite and arsenate from aqueous solutions [7]. Another sulfide material, pyrite FeS$_2$, has been studied with the goal to improve the solar energy applications [8] because it functioned as a photoelectrochemical solar cell, showing the stable behavior under illumination. The FeS$_2$ microspheres were analyzed by the scanning and transmission electron micrographs, the absorption spectra demonstrating the high absorbance in the visible region. These results indicate that FeS$_2$ is a promising candidate material for solar absorption application [9]. The ferromagnetic resonance was applied to study the natural Fe$_7$S$_8$ samples. In particular it was demonstrated that Pyrrhotite is not a good ferromagnetic resonance absorber, at least at the low frequencies [2]. The thermal expansivity of a natural single crystal of Pyrrhotite was measured, using the strain gauge technique. It was demonstrated that the anisotropy in this Pyrrhotite may come from the anisotropic stress field, produced by the temperature dependence of the elastic constants [9]. The ultrathin nanosheets of Pyrrhotite $Fe_7S_8$ with mixed valence states and metallic conductivity are demonstrated to be efficient for electrocatalysts of oxygen evolution reaction with highly active Fe sites [10].

It was found that the $Fe_7S_8$ compound [1,11] has a layered superstructure of the NiAs type with iron and sulfur layers alternatively. The planes holding sulfur atoms are complete. One-eighth of the iron sites are vacant. The superstructures appear from the ordering of vacancies in some iron layers. More specifically, the $Fe_7S_8$ compound shows two kinds of superstructures: the monoclinic one, 4C, and the hexagonal one, 3C. The structure of Pyrrhotite exhibiting a hexagonal 3C has been determined from analysis of three-dimensional single-crystal intensity data. This structure is composed of 12 sub-cell units stacked in three layers from four. The vacancies in the Fe sites are ordered and, confined to alternate layers of Fe atoms, normal to c [12]. At the same time, it was shown by neutron-diffraction measurements that the $Fe_7S_8$ possess a monoclinic 4C structure, similar to that in NiAs, in which cation completely occupies the layers alternating with the cation deficient layers [13]. Concerning the magnetic structure, there are ferromagnetically aligned layers with the antiferromagnetic coupling between adjacent layers. The electronic and magnetic structure of this lacunary sulfide is still under debates. For instance, it is not yet understood whether both $Fe^{3+}$ ions $Fe^{2+}$ ions are coexisting or only $Fe^{2+}$ ions are present. The X-ray magnetic circular dichroism at the Fe $L_{2,3}$ edges of $Fe_7S_8$ sample demonstrates, in coherence with multiplet calculations, that iron is present only as $Fe^{2+}$ [14]. At the same time, the results of X-ray photoelectron spectrometry demonstrate that $Fe_7S_8$ is composed of both $Fe^{2+}$ and $Fe^{3+}$ states [4].

In order to study theoretically magnetic properties and phase transitions of magnetic systems [15-21] many works have studied the mixed Ising model [22–24] by using different approximation methods like, Mean-Field Approximation (MFA) and Monte Carlo Simulation (MCS) [25–29].

In the present work, the structural, the electronic and the magnetic properties of $Fe_7S_8$ compound have been studied. In section II we define the theoretical model by using ab-initio method. To do this the framework of the Generalized Gradient Approximation (GGA), and the Full Potential Linearized Augmented Plane Wave (FP-LAPW) method implemented in the WIEN2K package are performed. In section III, the phase and ground states diagrams of the proposed mixed spin Ising model has been plotted and discussed. The details and numerical results of MFA are presented in section IV and V. MCS have been used to study the magnetic properties of $Fe_7S_8$ and the results have been discussed in section VI and VII. Finally, a conclusion is given in section VIII.

## II. Theoretical model and motivation
### II.1 Ab-initio calculations: Study of the density of state

Ab-initio calculations were used for investigation and analyzing the electronic DOS in the framework of the GGA. The FP-LAPW method was implemented within the WIEN2k package [30,31]. Up to 1000 k points were used to calculate the integrals in the irreducible Brillouin zone to ensure the optimal energy convergence [32,33]. The plane-wave cut-off was taken such a way to have $R_{MT}K_{max} = 7$ where $R_{MT}$ denotes the average radius of the muffin-tin spheres and $K_{max}$ signifies the maximal value of the wave vector, $K = k + G$.

Stability of the total energy of the system within $10^{-5}$ Ry was considered as the convergence criterium of the self-consistent calculations.

Our calculation are based from the structure of $Fe_7S_8$ proposed by Bertaut [11]. This structure is given by Figures 1-a and 1-b. The compound has a monoclinic structure 4C-type Pyrrhotite with the space group c2/c. The position of iron and sulfur atoms was taken as in the ideal NiAs-type structure. All the iron atoms lie between six sulfur atoms with octahedral arrangements and between other iron atoms. The sulfur atoms are coordinated by five iron atoms, which are located in the corners of a trigonal prism with one vacant site in one of the corners. Only sulfur S(4) is coordinated by six iron atoms.

a)                                              b)

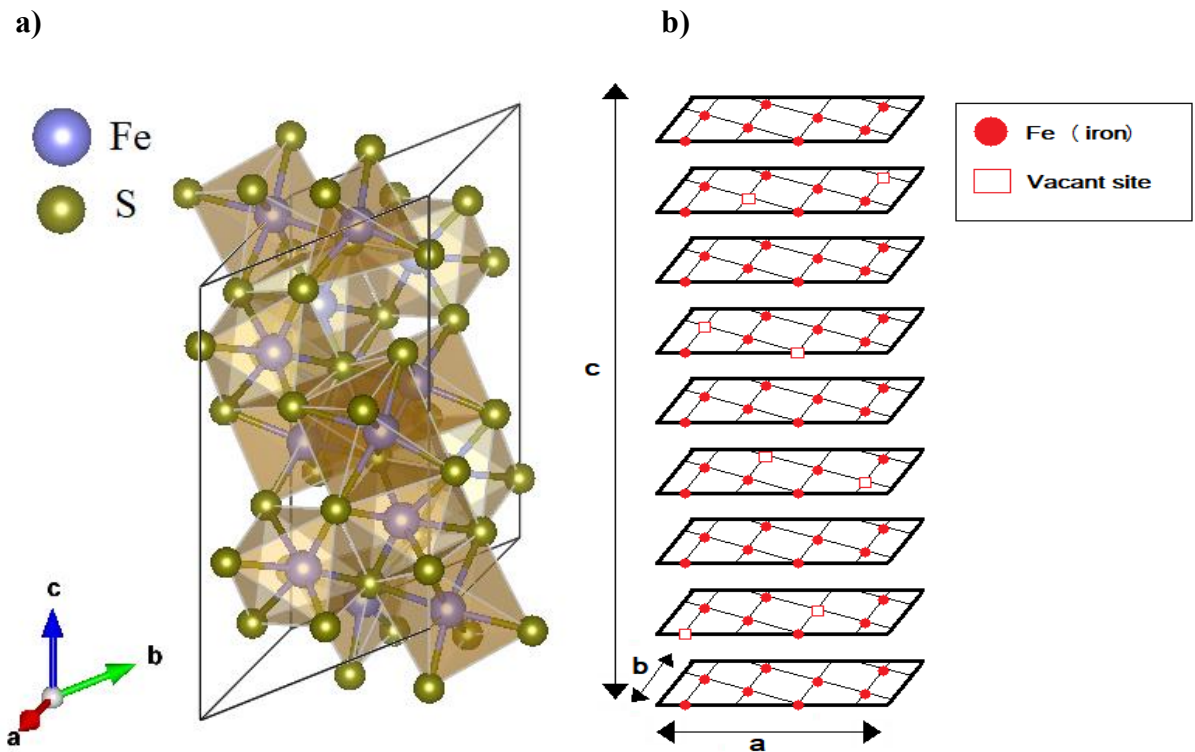

**Figure 1:** *a) The geometry of the crystal structure of the Monoclinic Pyrrhotite $Fe_7S_8$ and b) the magnetic structure.*

| Distances | |
|---|---|
| Fe(1)-S | 2.446 Å |
| Fe(2)-S | 2.449 Å |
| Fe(3)-S | 2.486 Å |
| Fe(4)-S | 2.444 Å |
| All Fe-Fe except Fe(3)-Fe(3) | 3.1 Å |
| Fe(3)-Fe(3) | 2.944 Å |
| Fe(4)-Fe(3) | 2.868 Å |
| Fe(2)-Fe(1) | 2.956 Å |
| Fe(3)-Fe(2) | 2.911 Å |

*Table 1: The distances of the ionic bonds.*

| Angles | |
|---|---|
| Fe(1)-Fe(2)-Fe(3) | 174.06° |
| Fe(2)-Fe(3)-Fe(1) | 166.05° |
| Fe(3)-Fe(1)-Fe(3) | 167.71° |

*Table 2: The angles of the ionic bonds.*

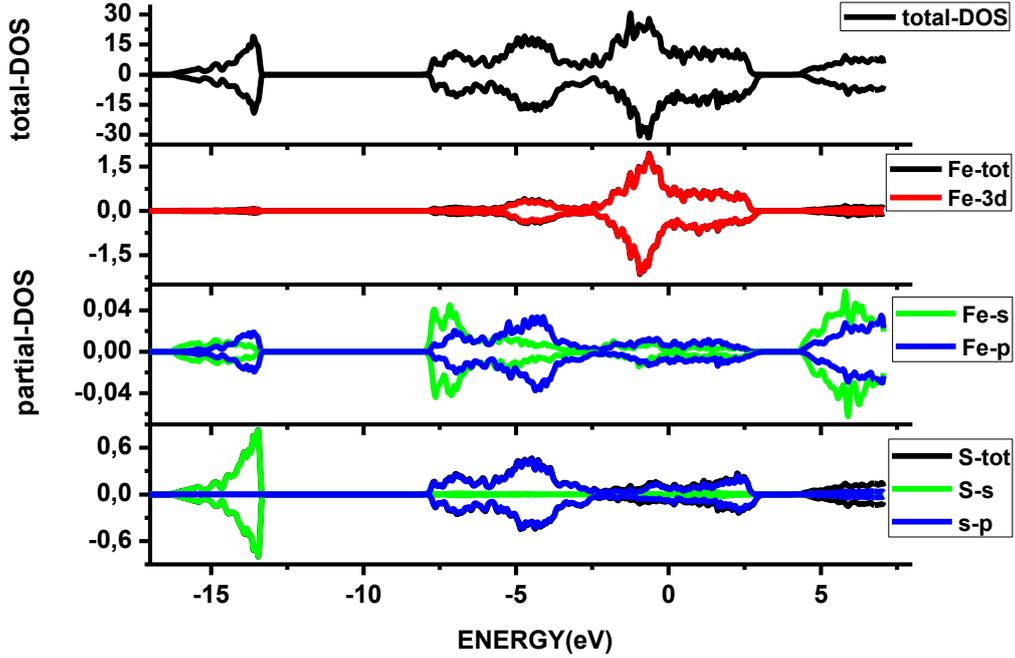

*Figure 2: The total and partial electronic densities of $Fe_7S_8$ compound.*

The DOS was calculated in order to investigate the nature of $Fe_7S_8$ system and the nature of the distribution of electrons in the valence and conduction bands. Figure 2 depicts the calculated total DOS of $Fe_7S_8$ and partial DOS for individual atoms Fe and S respectively, using GGA.

The valence band consists predominantly of Fe-3d and S-3p orbitals from -8 to 2 eV, while the S-3p is located from -17 eV to -12.5 eV. The conduction band is dominated by smaller contributions of S-3p states. Indeed, the Fe atoms are responsible for magnetism in the compound. The asymmetry of the spin states, "up" and "down" confirms that the structure is magnetic. The absence of the band gap proves the character metallic of $Fe_7S_8$ structure. The valence band contains dispersive states because the bands are very close and consequently electrons moved easily from one band to another. However, in the conduction band an electron must have a kinetic energy value for jumping from one band to another. The total magnetic moment is 3.970 $\mu_B$; this value is approximately equal to the magnetic moment given by ref. [13].

### II.2 The Hamiltonian

The compounds Fe ([Ar] $3d^6 4s^2$) and S ([Ne] $3s^2 3p^4$) constitute the structure of the $Fe_7S_8$ material (see Figure 1-a). From Wien2K's output files and looking at the on-site occupancies, as well as the spin states it shows that there are different oxidation states $\frac{2}{7}Fe^{3+}$, $\frac{5}{7}Fe^{2+}$ and $8S^{2-}$. The on-site occupancy and magnetization from DFT is better for distinguishing different oxidation states of 3d metals. The magnetic moment of this compound comes only from the $Fe^{2+}$ with spin value S = 2 and $Fe^{3+}$ with spin value σ = 5/2. According with the results of neutron diffraction [13] studies which showed that the magnetic moments of iron atoms are arranged ferromagnetically inside each layer, but coupled antiferromagnetically between adjacent layers.

Taking into account the three types of positions of Fe in this system, $Fe_1^{2+}$, $Fe_2^{2+}$ and $Fe^{3+}$, we have three sub-lattices, for which three different exchange interactions are considered. The recognition of the three sub-lattices is determined by the different number of nearest-neighbors between them in each sub-lattices. Also, the magnetic anisotropy of this material has been showed in our ab-initio calculations.

Thus, in order to study the phase transition and the magnetic properties of the $Fe_{1-x}S$ like we use the mixed spin Ising model [34–40] with crystal field parameter as an appropriate energy. Then our system can be described by the following Hamiltonian:

$$H = -J_{S_1-\sigma} \sum_{<ik>} S_{1i}\sigma_k - J_\perp \sum_{<ij>} S_{1i}S_{2j} - J_{\sigma-\sigma} \sum_{<kl>} \sigma_k\sigma_l - \Delta_s \sum_i (S_i)^2 - \Delta_\sigma \sum_k (\sigma_k)^2 \quad (1)$$

Where $S_{1i}$, $S_{2j}$ and $\sigma_k$ are spins of the three sub-lattices of the system, $\sum_{<ij>}$ denotes the sum over nearest neighbors of i and j sites and $J_{S_1-\sigma} > 0$ is the ferromagnetic exchange interaction between the spins $\sigma$ and $S_1$ that are coming from $Fe^{2+} - Fe^{3+}$ interaction. $J_\perp$ is the antiferromagnetic exchange interaction between $S_1$ and $S_2$ spins of $Fe^{2+} - Fe^{2+}$ coming from adjacent layers and $J_{\sigma-\sigma} > 0$ is the ferromagnetic $\sigma - \sigma$ spin exchange interaction of $Fe^{3+} - Fe^{3+}$. The parameters $\Delta_s$ and $\Delta_\sigma$ are the crystal fields on sites $Fe^{2+}$ and $Fe^{3+}$ respectively.

The crystal field is the most widely adopted form in quantum spin models to describe the anisotropies in magnetic systems like for example in ref. [41,42]. In our calculation, different values of $\Delta_s$ and $\Delta_\sigma$ are chosen in order to see the effect of the crystal field on the system properties. That is because, the negative crystal field gives us the smaller state of spin and the positive one gives the bigger values of spin.

### III. Phase diagrams and ground state of the system at $T/K_B J_{\sigma-\sigma}=0$

At temperature $T/K_B J_{\sigma-\sigma}=0$, the ground states of the system described by the Hamiltonian (1) are obtained by the computation and comparison of energies of all possible phases of $(S_1, S_2, \sigma)$. The possible spin values of $\sigma$ can take the spin values of $\pm 5/2$, $\pm 3/2$, $\pm 1/2$, and these of $S_1$ and $S_2$ are $\pm 2$, $\pm 1$, $0$. The results of calculations reveal that there are 17 stable phases given by the Table 3 below:

| Code | Corresponding spin configuration of $(S_1, S_2, \sigma)$ |
|---|---|
| A | (2, -2, 5/2) |
| B | (2, 2, 5/2) |
| C | (2, -2, -5/2) |
| D | (2, 2, -5/2) |
| G | (2, -1, -5/2) |
| I | (2, 0, -5/2) |
| J | (2, -1, 5/2) |
| K | (2, 0, 5/2) |
| L | (2, 1, 5/2) |
| M | (1, 1, 5/2) |
| N | (1, 1, -5/2) |
| Brown | (0, 0, 5/2) |
| P | (2, 1, -5/2) |
| Yellow | (1, -1, 5/2) |
| Green | (1, -1, -5/2) |
| Blue | (1, 0, 5/2) |
| Red | (1, 0, -5/2) |

***Table 3:*** *Code of each configuration.*

The phase diagrams, given in Figures 3-a and 3-b, show stability regions for each $(S_1, S_2, \sigma)$ phase in the $(J_\perp, J_{S_1-\sigma})$ plane. Figures 3 are performed for $J_{\sigma-\sigma} = 1$ as an arbitrary values and for $\Delta_s = \pm 1$ and $\Delta_\sigma=1$.

Four phases are found in figure 3-a: three ferromagnetic phases A, C and D, and ferrimagnetic one B. The phases are separating by coexistence lines with the same phases energies. The critical point $\Gamma_1$ ($J_\perp=0, J_{S_1-\sigma} = 0$) separates two regions: for $J_\perp < 0$ three phase transitions can be found (from D to C), (from C to A) and (from A to B), when $J_{S_1-\sigma}$ increases. For $J_\perp > 0$ only one phase transition D to B is possible. However, in the critical point $\Gamma_1$ the four phases (A, B, C and D) coexist.

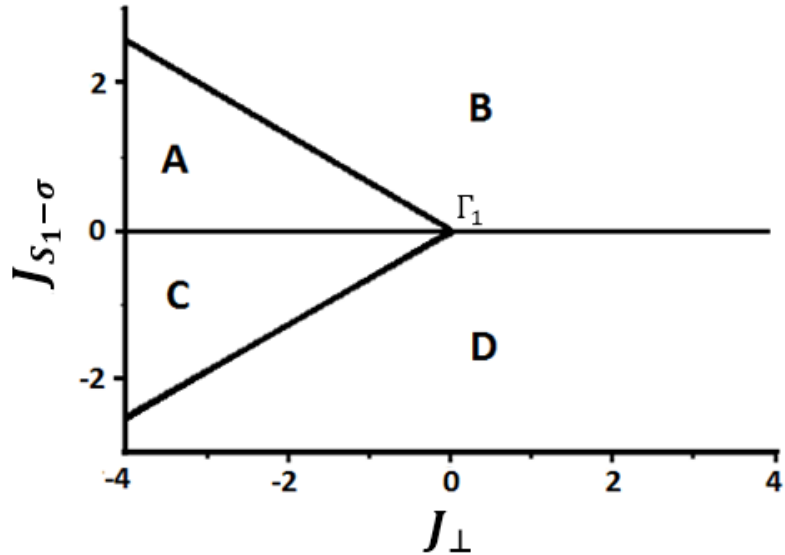

**Figure 3-a:** *The phase diagram of the system for $\Delta_s = \Delta_\sigma = 1$ and $J_{\sigma-\sigma} = 1$.*

Figure 3-b shows seventeen stable phases: Eleven phases are ferrimagnetic A, C, D, G, I, P, J, N, Yellow, Green and Red, and six ferromagnetic phases B, K, L, M, Brown and Blue. Like in Figure 3-a, the phases are separating by coexistence lines with the same phases energies. We observe two critical points: $\Gamma_2$ at ($J_\perp = -3.5, J_{S_1-\sigma} = 0$) where five phases are coexisting (A, C Yellow, Brown and Green), and $\Gamma_3$ at ($J_\perp = 3, J_{S_1-\sigma} = 0$) where three phases are coexisting (B, D and Brown). When $J_{S_1-\sigma} > 0.1$ there are eight phases (A, J, K, L, B, M, Blue and Yellow). When $J_{S_1-\sigma} < -0.1$ there are eight ferrimagnetic phases (C, G, I, P, D, N, Red and Green). For $-0.1 < J_{S_1-\sigma} < 0.1$ and $-3 < J_\perp < 3$, the stable ferromagnetic phase is Brown.

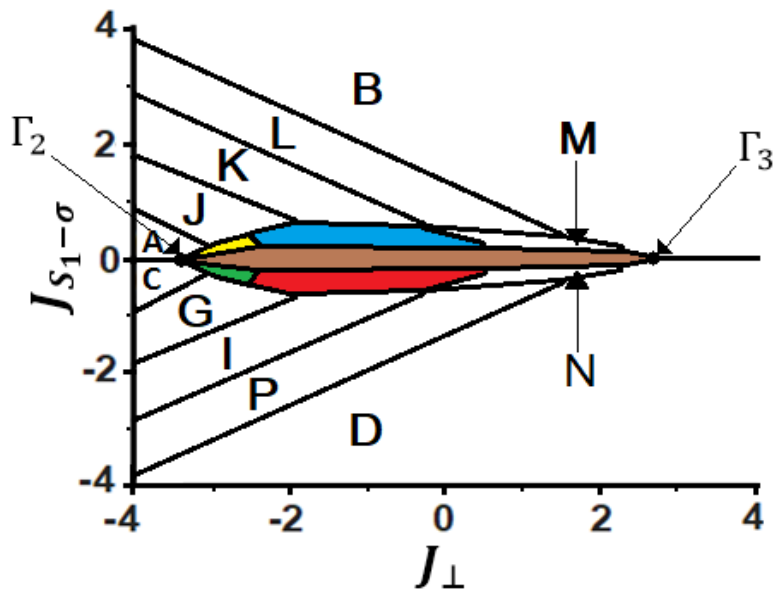

**Figure 3-b:** *The phase diagram of the system for $\Delta_s = -1, \Delta_\sigma = 1$ and $J_{\sigma-\sigma} = 1$.*

The mean field approximation was used to explore the qualitative magnetic behavior of the studied system as a function of temperature and compare with the results of the previous Section. The variational method, in which the correlations between spins were neglected was based on the Gibbs-Bogoliubov inequality for the exact free energy F per site of N-body system [43,44]:

$$F \leq F_0$$

Where

$$F_0 = -K_B T \ln(Z_0) + \langle H - H_0 \rangle_0$$

$F_0$ is the approached free energy which depends on the effective Hamiltonian $H_0$, $Z_0$ corresponds to the approached partition function, T is the absolute temperature and $H_0$ denotes the effective Hamiltonian in which the interactions between each spin with its neighbors are substituted by an effective field h. According to (1) this effective Hamiltonian can be decomposed as:

$$H_0 = H_0^{S1} + H_0^{S2} + H_0^{\sigma}$$

Where $H_0^{S1}$, $H_0^{S2}$ and $H_0^{\sigma}$ are the effective Hamiltonians corresponding to the three sub lattices of the model:

$$H_0^{S1} = -h_1^{S1} \sum_i S_{1i} - h_2^{S1} \sum_i S_{1i} - \Delta_s \sum_i (S_{1i})^2,$$

$$H_0^{S2} = -h_1^{S2} \sum_i S_{2i} - h_2^{S2} \sum_i S_{2i} - \Delta_s \sum_i (S_{2i})^2$$

and

$$H_0^{\sigma} = -h_1^{\sigma} \sum_k \sigma_k - h_2^{\sigma} \sum_k \sigma_k - \Delta_\sigma \sum_k (\sigma_k)^2.$$

Where $h_1^{S1} = z_1 J_{S1-\sigma} \langle \sigma \rangle$, $h_2^{S1} = z_2 J_\perp \langle S_2 \rangle$, $h_1^{S2} = z_1 J_{S1-\sigma} \langle \sigma \rangle$, $h_2^{S2} = z_2 J_\perp \langle S1 \rangle$, $h_1^\sigma = z_3 J_{S1-\sigma} \langle S1 \rangle$ and $h_2^\sigma = z_4 J_{\sigma-\sigma} \langle \sigma \rangle$.

Where $M_\sigma = \langle \sigma \rangle / \mu_B$ is the $\sigma$ sub-lattice magnetization, $M_{S2} = \langle S_2 \rangle / \mu_B$ is the $S_2$ sub-lattice magnetization and $M_{S1} = \langle S_1 \rangle / \mu_B$ is the $S_1$ sub-lattice magnetization. All magnetizations have been normalized by Bohr magneton $\mu_B$. $z_1 = 4$ is the first nearest neighbors $\sigma$ of spin site $S_1$, $z_2 = 2$ is the first nearest neighbors $S_2$ of spin site $S_1$, $z_3 = 3$ is the first nearest neighbors $S_1$ of spin site $\sigma$ and $z_4 = 2$ is the first nearest neighbors $\sigma$ of spin site $\sigma$.

For each sub-lattice we used the effective field in the simplified form, given by:
$h_I = h_1^{S1} + h_2^{S1}$, $h_{II} = h_1^{S2} + h_2^{S2}$ and $h_{III} = h_1^\sigma + h_2^\sigma$.
After calculation of the approached partition function $Z_0 = \text{Tr} e^{-\beta H_0}$, and after minimization the free energy $F_0$ by using the relationships:
$M_{S1} = -\frac{\partial F_0}{\partial h_I}$, $M_{S2} = -\frac{\partial F_0}{\partial h_{II}}$ and $M_\sigma = -\frac{\partial F_0}{\partial h_{III}}$.

We obtain the sub-lattice magnetizations:

$$M_{S1} = \frac{4 \sinh\left(\frac{2h_I}{K_B T}\right) e^{\left(\frac{4\Delta_s}{K_B T}\right)} + 2 \sinh\left(\frac{h_I}{K_B T}\right) e^{\left(\frac{\Delta_s}{K_B T}\right)}}{1 + 2 \sinh\left(\frac{2h_I}{K_B T}\right) e^{\left(\frac{4\Delta_s}{K_B T}\right)} + 2 \cosh\left(\frac{h_I}{K_B T}\right) e^{\left(\frac{\Delta_s}{K_B T}\right)}},$$

$$M_{S2} = \frac{4 \sinh\left(\frac{2h_{II}}{K_B T}\right) e^{\left(\frac{4\Delta_s}{K_B T}\right)} + 2 \sinh\left(\frac{h_{II}}{K_B T}\right) e^{\left(\frac{\Delta_s}{K_B T}\right)}}{1 + 2 \sinh\left(\frac{2h_{II}}{K_B T}\right) e^{\left(\frac{4\Delta_s}{K_B T}\right)} + 2 \cosh\left(\frac{h_{II}}{K_B T}\right) e^{\left(\frac{\Delta_s}{K_B T}\right)}}$$

and

$$M_\sigma = \frac{5 \sinh\left(\frac{5h_{III}}{2K_B T}\right) e^{\left(\frac{25\Delta_\sigma}{4K_B T}\right)} + 3 \sinh\left(\frac{3h_{III}}{2K_B T}\right) e^{\left(\frac{9\Delta_\sigma}{4K_B T}\right)} + \sinh\left(\frac{h_{III}}{2K_B T}\right) e^{\left(\frac{\Delta_\sigma}{4K_B T}\right)}}{2 \cosh\left(\frac{5h_{III}}{2K_B T}\right) e^{\left(\frac{25\Delta_\sigma}{4K_B T}\right)} + 2 \cosh\left(\frac{3h_{III}}{2K_B T}\right) e^{\left(\frac{9\Delta_\sigma}{4K_B T}\right)} + 2 \cosh\left(\frac{h_{III}}{2K_B T}\right) e^{\left(\frac{\Delta_\sigma}{4K_B T}\right)}}.$$

Where $K_B$ is the Boltzmann constant.
Finally, the total magnetization of the system is given by
$$M_T = M_{S1} + M_{S2} + M_\sigma$$

## IV. Mean Field results: Finite-temperature phase diagrams and magnetic properties
### IV.1 Case where $\Delta_s = \Delta_\sigma = 1$, $J_{S1-\sigma} = 1$ and $J_{\sigma-\sigma} = 1$

Figure 4 shows the critical temperature (blue line) and first order phase transitions of the system (dashed line) in $(J_\perp, T/K_B J_{\sigma-\sigma})$ plane, only for the part where $-4 < J_\perp < 0$ (see Figure 3-a). Accordance to diagram of Figure 3-a, for $J_{S_1-\sigma} = 1$, only two magnetic phases (2, 2, 5/2) and (2, -2, 5/2) are stable for $(T/K_B J_{\sigma-\sigma} = 0)$. For finite temperatures a magnetic phase (-2, -2, 3/2) appears below the critical temperature line. An end point $(J_\perp = -2.5, T/K_B J_{\sigma-\sigma} = 0.5)$ can be observed where first and second phase transition coexist.

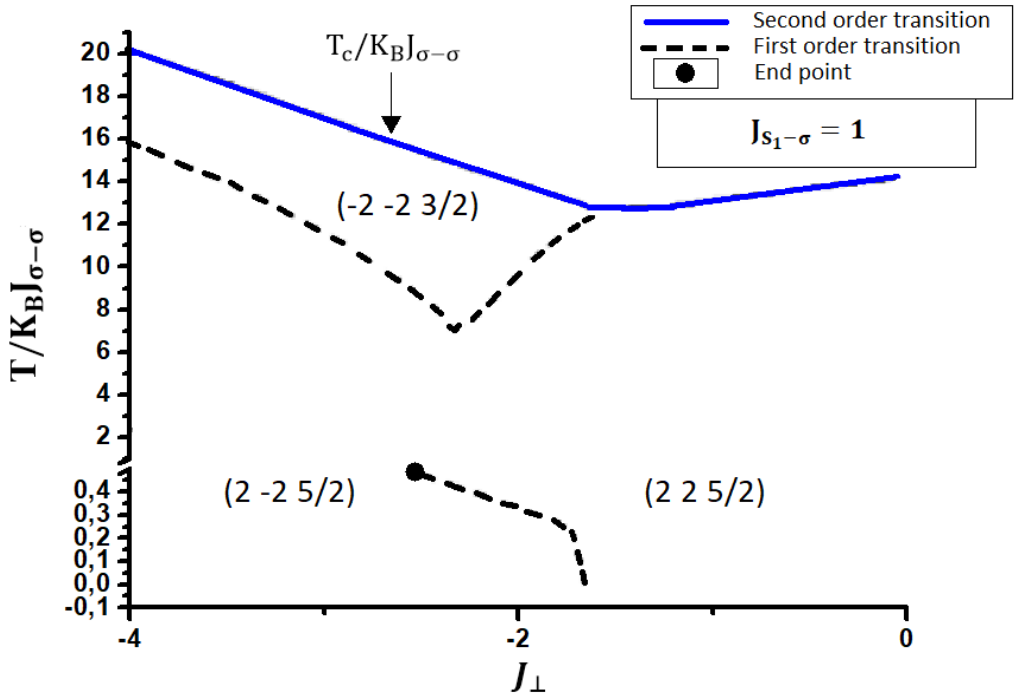

*Figure 4: Phase diagram transition.*

The behavior of $T_c/K_B J_{\sigma-\sigma}$ can be explained by the competition between the ferromagnetic interactions ($J_{S_1-\sigma}$ and $J_{\sigma-\sigma}$) and the antiferromagnetic one $J_\perp$. Table 4 gives a summary of Figure 4 where the first order transition and second order ones are given for different interval values of $J_\perp$.

| $J_\perp$ interval values | *First order transition* | *Second order transition* |
|---|---|---|
| $J_\perp < -2.5$ | $\left(2, -2, \frac{5}{2}\right) \to \left(-2, -2, \frac{3}{2}\right)$ | $\left(-2, -2, \frac{3}{2}\right) \to$ paramagnetic phase |
| $-2.5 < J_\perp < -1.6$ | $\left(2, 2, \frac{5}{2}\right) \to \left(2, -2, \frac{5}{2}\right)$<br>From $T/K_B J_{\sigma-\sigma}=0$ to end point | None |
| $J_\perp > -1.55$ | $\left(2, 2, \frac{5}{2}\right) \to \left(-2, -2, \frac{3}{2}\right)$ | $\left(-2, -2, \frac{3}{2}\right) \to$ paramagnetic phase |

*Table 4: Summary phase transitions of Figure 4.*

Figures 5-a, 5-b and 5-c show the total and sub-lattice magnetizations for several values of $J_\perp$. These Figures are plotted in two interesting regions of Figure 4: a region $J_\perp < -1.6$, where $T_c/K_B J_{\sigma-\sigma}$ decreases and a region $J_\perp > -1.6$, where $T_c/K_B J_{\sigma-\sigma}$ increases.

In Figure 5-a, the strength of the antiferromagnetic coupling, $J_\perp$, is not important enough to observe the first order transition of the all magnetizations, unlike Figures 5-b and 5-c.

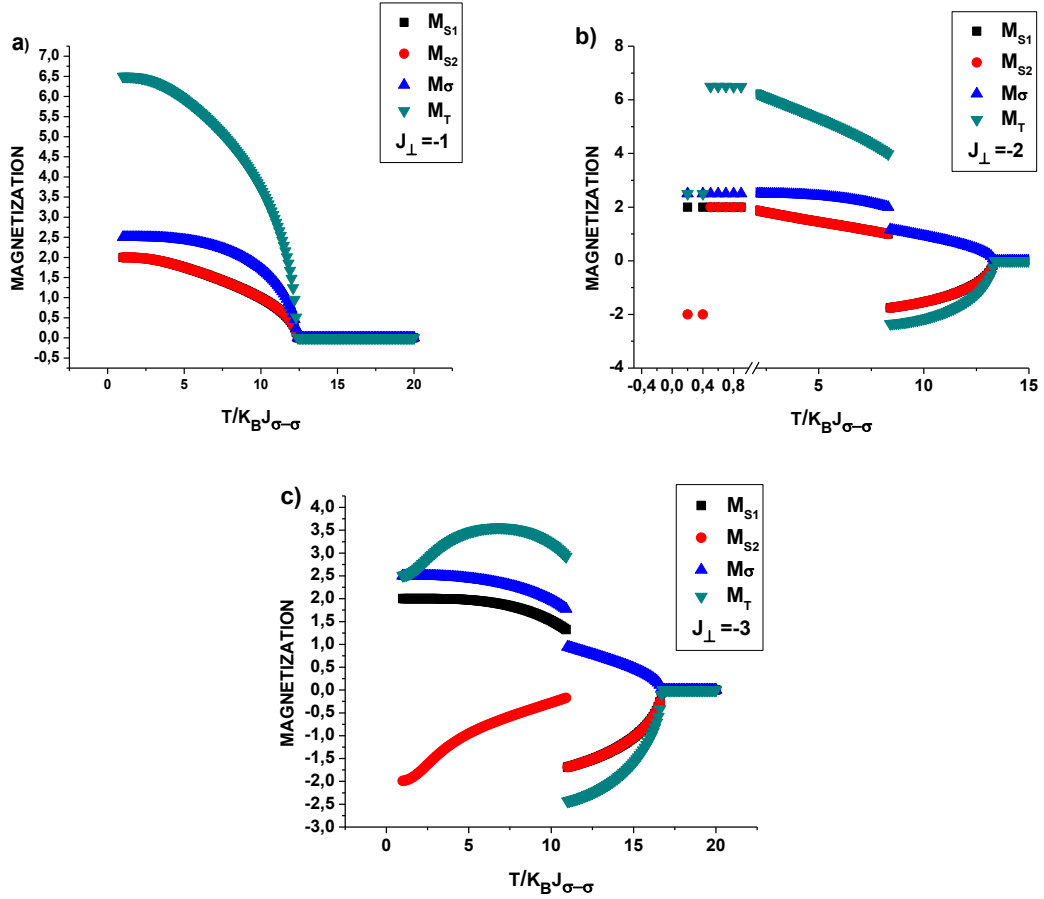

*Figure 5:* *The total and sub-lattice magnetizations.*

### IV.2     Case where $\Delta_s = -1, \Delta_\sigma = 1, J_{S1-\sigma} = 1$ and $J_{\sigma-\sigma} = 1$

Figure 6 shows the critical temperature (blue line) and first order phase transitions of the system (dashed line) in $(J_\perp, T/K_B J_{\sigma-\sigma})$ plane, only for the part where $-4 < J_\perp < 0$ of Figure 3-b. At $T/K_B J_{\sigma-\sigma} < 5$ only five magnetic phases are stable namely (2, -2, 5/2), (2, -1, 5/2), (2, 0, 5/2), (2, 1, 5/2) and (2, 2, 5/2). An another magnetic phase (2, 2, -3/2) appears at $T/K_B J_{\sigma-\sigma} > 5$ on below the critical temperature line. Four end points $(J_\perp = -3.3, T/K_B J_{\sigma-\sigma} = 4), (J_\perp = -2.8, T/K_B J_{\sigma-\sigma} = 4.2), (J_\perp = -2.7, T/K_B J_{\sigma-\sigma} = 4.4)$ and $(J_\perp = -2.5, T/K_B J_{\sigma-\sigma} = 4)$ can be observed where first and second phase transition coexist.

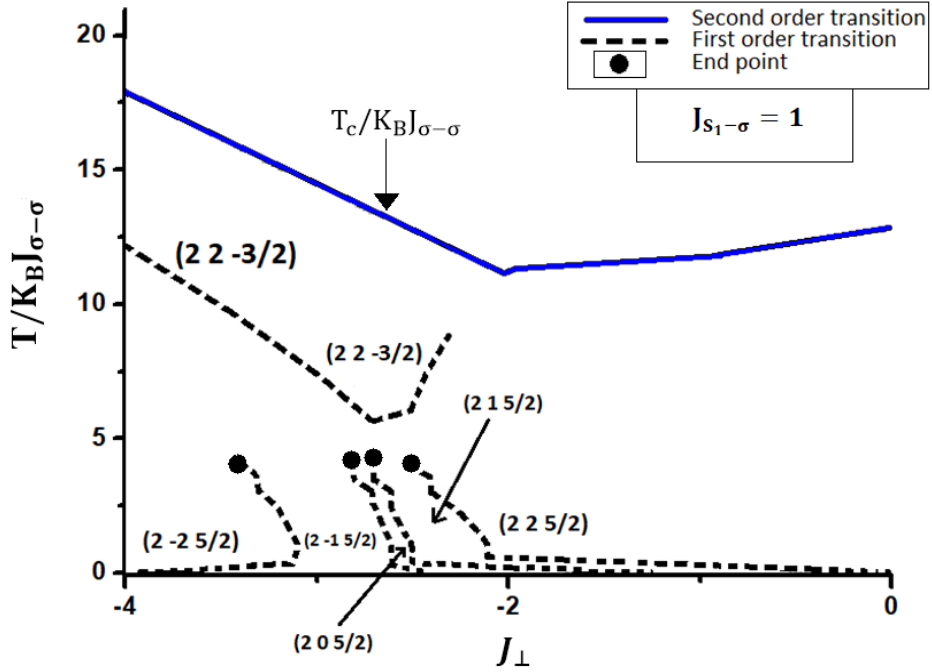

*Figure 6: Phase diagram transition.*

The behavior of $T_c/K_B J_{\sigma-\sigma}$ can be explained by the competition between the ferromagnetic interactions ($J_{S_1-\sigma}$ and $J_{\sigma-\sigma}$) and the antiferromagnetic one $J_\perp$.

Figures 7-a, 7-b, 7-c and 7-d show the total and sub-lattice magnetizations for several values of $J_\perp$. These Figures are plotted in two interesting regions of Figure 6: a region $J_\perp < -2.1$, where $T_c/K_B J_{\sigma-\sigma}$ decreases and a region $J_\perp > -2.1$, where $T_c/K_B J_{\sigma-\sigma}$ increases.

The first order transition of the sub-lattice magnetizations can be observed in Figures 7-a, 7-b and 7-c, unlike Figure 7-d because the $J_\perp$, is important.

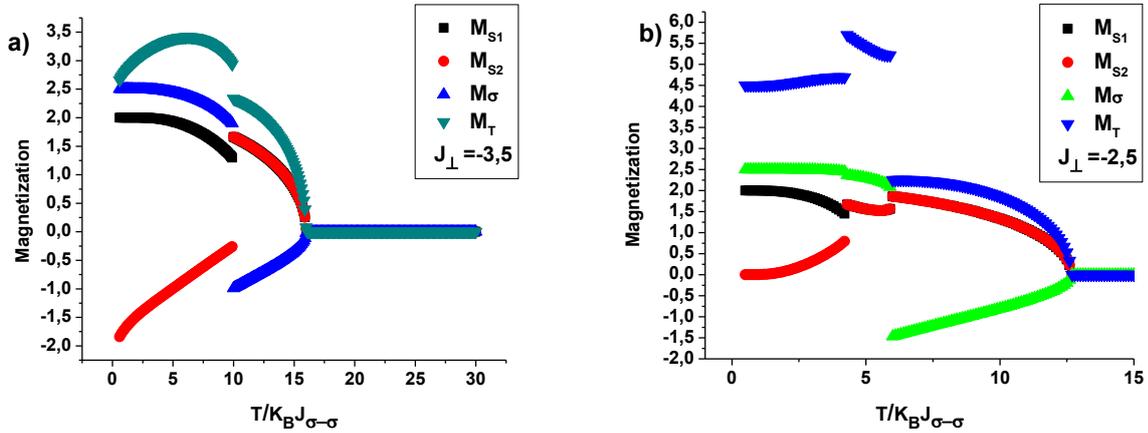

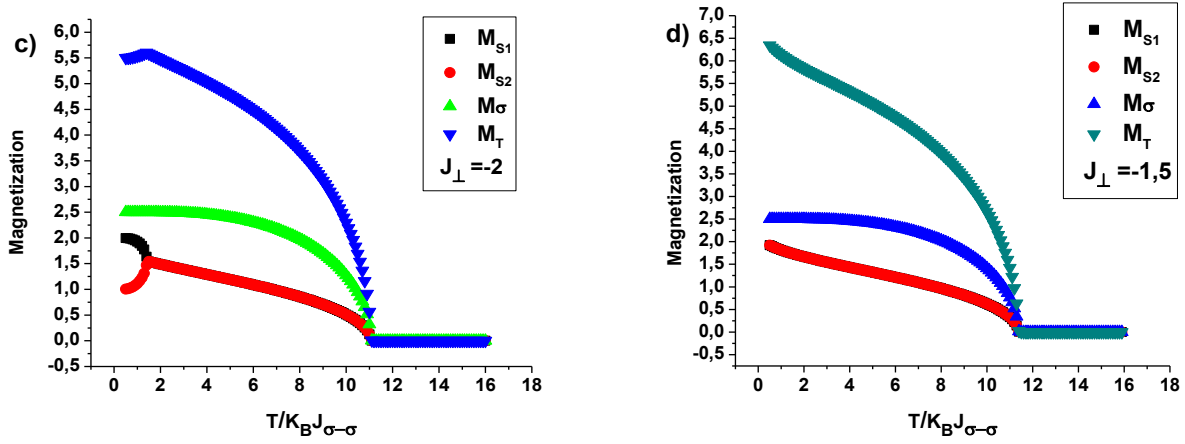

*Figure 7: The total and sub-lattice magnetizations.*

Compared with the experimental studies given in Figure 8-a of ref. [1], the increasing behavior of the total magnetization M$_T$ at low temperature is observed by our calculations in Figure 7-c. This anomaly can be explained by the non-uniform variation of the sub-lattices magnetizations. The difference between the experimental Neel 590K of Figure 8-a and the calculated temperature transition of our study is due to the fact that the temperature is normalized by $K_B J_{\sigma-\sigma}$.

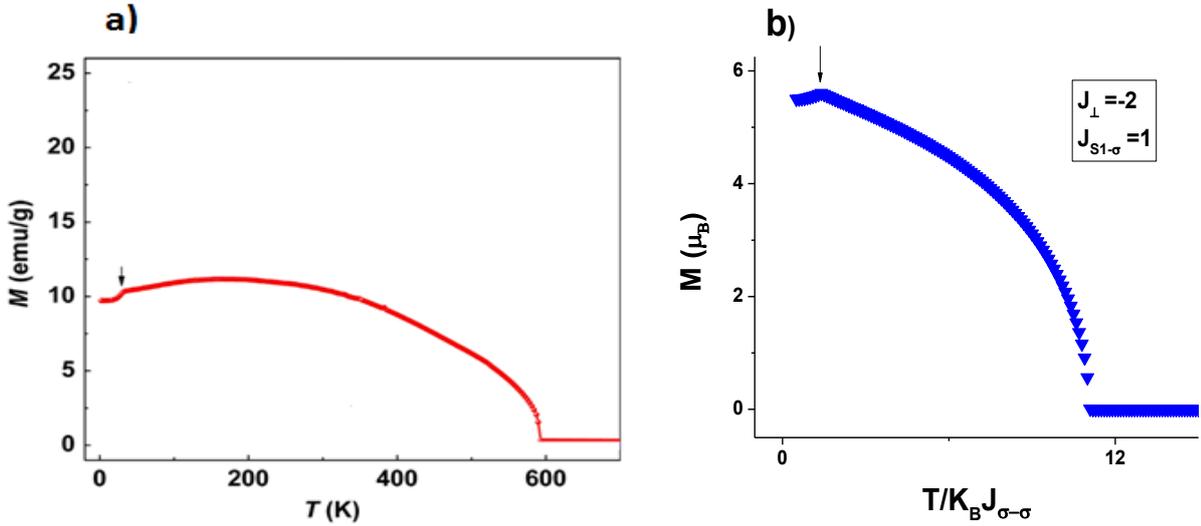

*Figure 8: a) Experimental results of the magnetization [1], and b) total magnetization of our calculations.*

## V. Monte Carlo simulations

In order to perform the Monte Carlo calculation of the Hamiltonian (1) and the geometry presented in Figure 1-b, the total and sub-lattice magnetizations, magnetic susceptibility and specific heat are calculated using the following equations respectively:

$$M_{S_1} = \frac{1}{N^3} \langle \sum_i S_{1i} \rangle,$$

$$M_{S_2} = \frac{1}{N^3} \langle \sum_i S_{2i} \rangle,$$

$$M_\sigma = \frac{1}{N^3} \langle \sum_i \sigma_i \rangle,$$

$$M_T = M_{S_1} + M_{S_2} + M_\sigma,$$

$$\chi = \frac{N}{K_B T}\big(\langle M_T^2 \rangle - \langle M_T \rangle^2\big),$$

and

$$C_v = \frac{1}{N(K_B T)^2}\big(\langle E^2 \rangle - \langle E \rangle^2\big).$$

Where E is the internal energy of the system and N is the number of spin in each sub-lattice.

The following analysis will be based on the periodic boundary conditions. The $10^5$ Monte Carlo steps are used for each spin configuration.

## VI. Monte Carlo results: Finite-temperature magnetic properties
### VI.1 Case where $\Delta_s = \Delta_\sigma = 1$, $J_{S1-\sigma} = 1$ and $J_{\sigma-\sigma} = 1$

In Figures 9 and 10, the total magnetic susceptibility and the total magnetizations and also the sub-lattice magnetizations are given as function of the temperatures for $J_\perp = -2$ and $J_\perp = -3$ respectively.

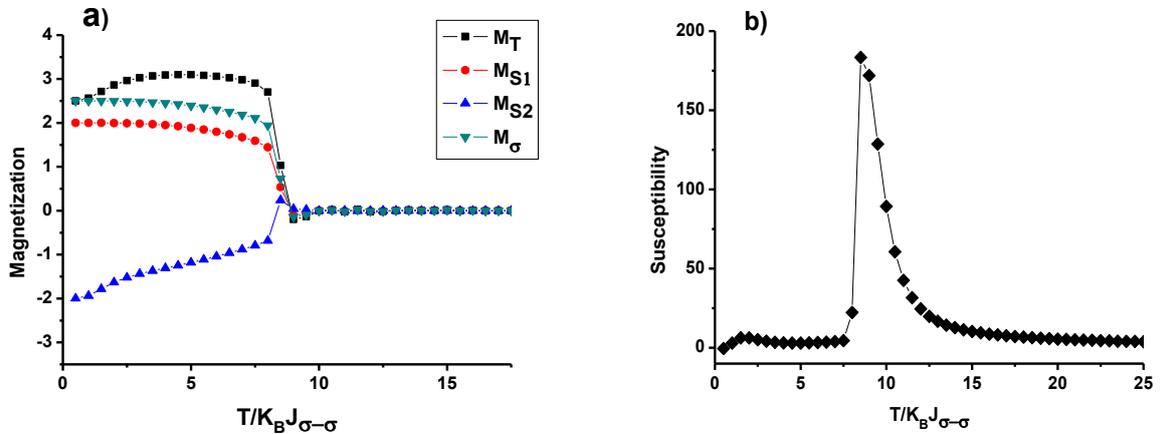

***Figure 9:*** *a) Total and sub-lattice magnetizations and b) total magnetic susceptibility for $J_\perp = -2$.*

In Figure 9-a, due to the negative value of the sub-lattice magnetization $M_{S2}$, we can see the same increasing behavior of the total magnetization at low temperature, observed by MFA of Figure 8-b and by the experimental study of Figure 8-a. Both sub-lattice magnetizations and total magnetization vanish at $T_c/K_B J_{\sigma-\sigma}=8.5$. This value of $T_c/K_B J_{\sigma-\sigma}$ corresponds to the peak of the total magnetic susceptibility of Figure 9-b.

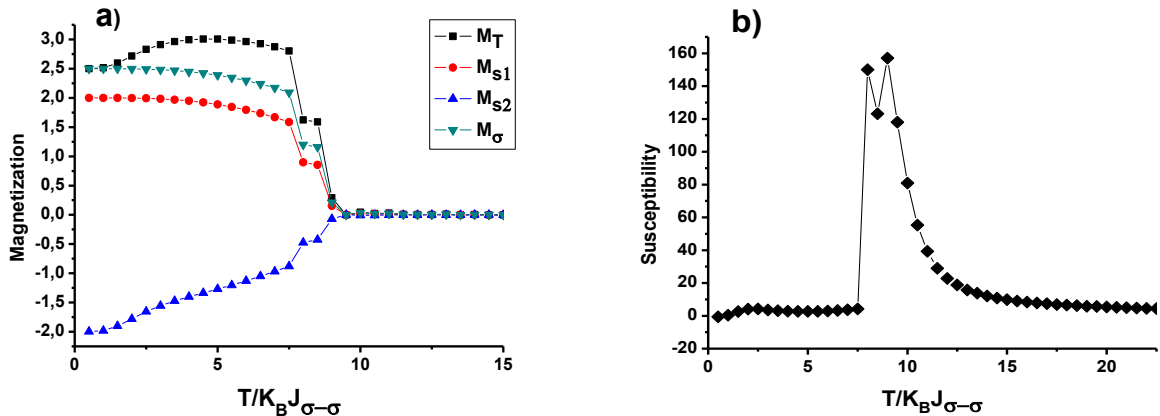

***Figure 10:*** *a) Total and sub-lattice magnetizations and b) total magnetic susceptibility for $J_\perp = -3$.*

In contrary of Figure 9-a, the first order transition of all the magnetizations have been observed in Figure 10-a. The critical temperature $T_c/K_B J_{\sigma-\sigma}=9$ is given by the second peak of total magnetic susceptibility of Figure 10-b. However the first peak of Figure 10-b is due to the first order transition of the magnetizations observed below $T_c/K_B J_{\sigma-\sigma}$.

Once again the MCS show the same increasing behavior of the total magnetization at low temperature observed by MFA (Figure 8-b) and experimental study (Figure 8-a).

From the total susceptibility magnetic of Figures 9-b and 10-b, we can conclude that when the antiferromagnetic coupling increases, for a fixed value of the ferromagnetic coupling, the transition temperature $T_c/K_B J_{\sigma-\sigma}$ decreases.

### VI.2    Case where $\Delta_S = -1, \Delta_\sigma = 1$ and $J_{\sigma-\sigma} = 1$

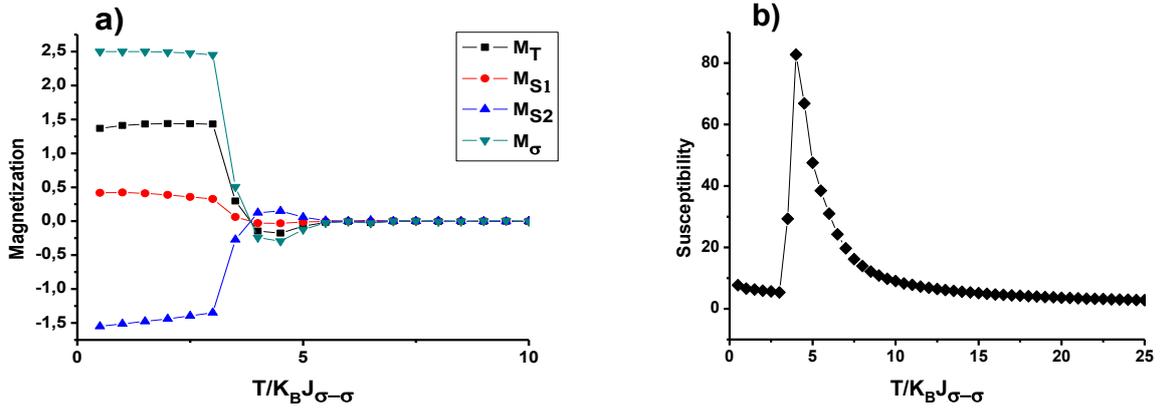

*Figure 11:* a) Total and sub-lattice magnetizations and b) total magnetic susceptibility for $J_{S1-\sigma} = 0.1$ and $J_\perp = -2.7$.

In contrary to MFA results, the Figure 11-a shows a compensation temperature at $T/K_B J_{\sigma-\sigma} = 3.5$ below the second order phase transition which occurs at $T_c/K_B J_{\sigma-\sigma} = 4$. The value of critical temperature $T_c/K_B J_{\sigma-\sigma}$ corresponds to the peak of the total magnetic susceptibility of Figure 11-b. One can also conclude that the negative value of the crystal field ($\Delta_S = -1$) favors the phenomena of compassing in this system.

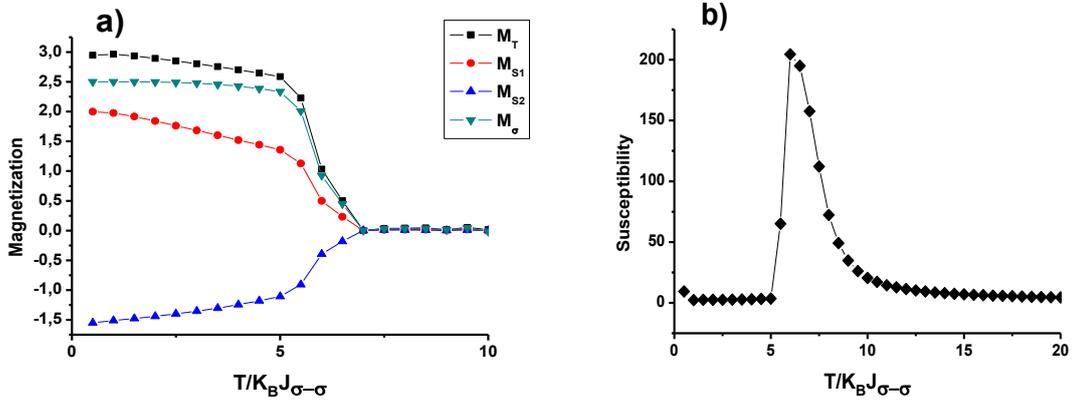

*Figure 12:* a) Total and sub-lattice magnetizations and b) total magnetic susceptibility for $J_{S1-\sigma} = 0.7$ and $J_\perp = -2.7$.

In Figure 12-a, the compensation temperature disappears. This can be explain by the important value of $J_{S1-\sigma}$ contrary to that of Figure 11-a. This because that the ferromagnetic sub-lattices coupling are stronger than the antiferromagnetic one. The second order transition of the magnetizations occurs at the critical temperature $T_c/K_B J_{\sigma-\sigma} = 7$, as we can see by the peak of the total magnetic susceptibility Figure 12-b.

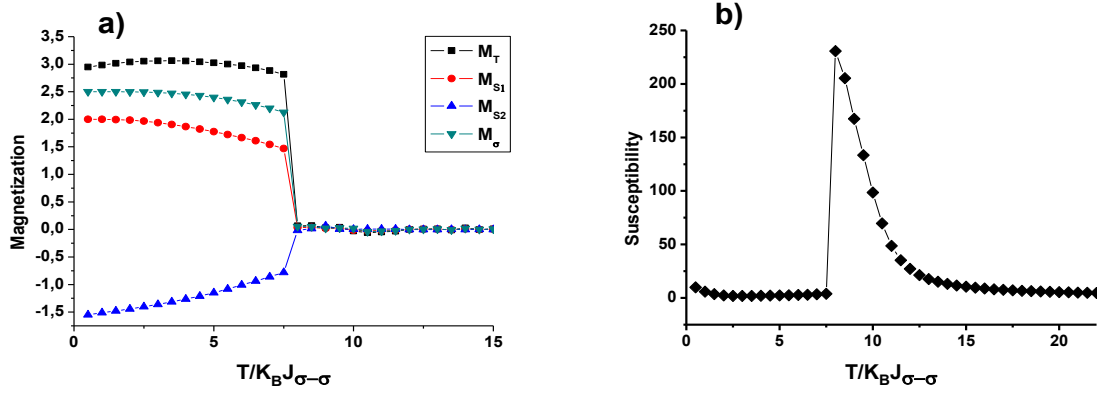

*Figure 13: a) Total and sub-lattice magnetizations and b) total magnetic susceptibility for $J_{S1-\sigma} = 1.2$ and $J_\perp = -2.7$.*

In Figures 13-a and 14-a, a rapid decreasing of the total magnetization curves can be observed around $T_c/K_B J_{\sigma-\sigma} = 7.5$. This behavior is due to the important competition between ferromagnetic and antiferromagnetic, $J_{S1-\sigma}$ and $J_\perp$ respectively.

From the total susceptibility magnetic of Figures 11-b, 12-b and 13-b, we can conclude that when the ferromagnetic coupling increases, for a fixed value of the antiferromagnetic coupling, the transition temperature $T_c/K_B J_{\sigma-\sigma}$ increases.

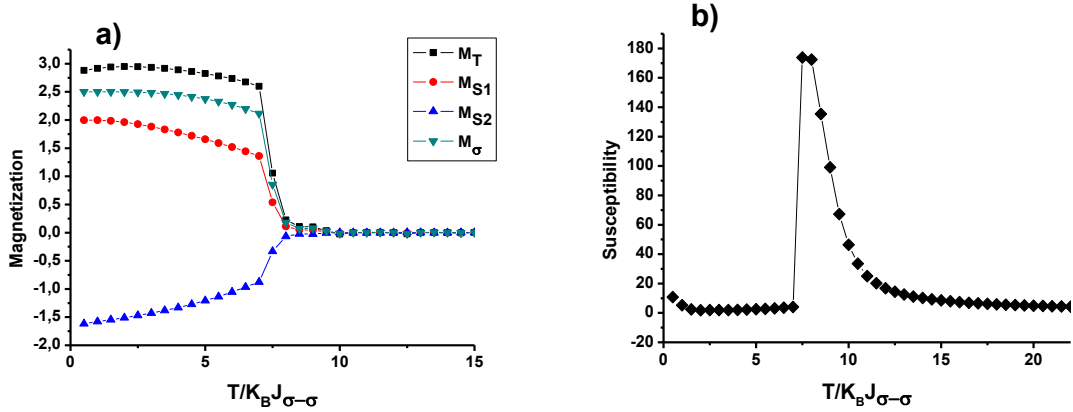

*Figure 14: a) Total and sub-lattice magnetizations and b) total magnetic susceptibility for $J_{S1-\sigma} = 1$ and $J_\perp = -3.5$.*

## VII. Conclusion

In this work, the structural and electronic properties of $Fe_7S_8$ has been studied using ab-initio calculations. It is found that the valence band consists, predominantly, of Fe-3d and S-3p. From Wien2K's output files it shows that there are different oxidation states namely $\frac{2}{7}Fe^{3+}$, $\frac{5}{7}Fe^{2+}$ and $8S^{2-}$. Thus the magnetic moment of this compound comes only from the $Fe^{2+}$ with spin value S = 2 and $Fe^{3+}$ with spin value σ = 5/2. The magnetic moments of iron atoms are arranged ferromagnetically inside each layer, but coupled antiferromagnetically between adjacent layers. The ab-initio calculations show also a presence of the magnetic anisotropy in this material.

Taking into account the results of ab-initio calculations, a mixed Ising model with crystal field has been chosen in order to describe the magnetic properties and the phase transition of $Fe_7S_8$ compound. The ground state phases, at $T/K_B J_{\sigma-\sigma}$=0, of this model are analyzed for different exchange couplings and for different values of the crystal fields. Depending on the exchange interactions of this model many stable phases can be found namely ferromagnetic and ferrimagnetic ones.

By using MFA and MCS, it is found that the system present the first order transitions of the sub-lattices magnetizations below the critical temperature of the system. And as it is found by the experiment study of ref. [1], the both methods show an anomaly of increasing behavior of the total magnetization at the low temperature. This anomaly can be explained by the non-uniform variation of the sub-lattices magnetizations on below the critical temperature.

It is found also that a compensation temperature exists in MCS but it is not the case for the MFA results.

The total magnetic susceptibility are calculated for various values of exchange interactions and crystal fields parameters. The results have shown that when the ferromagnetic coupling increases, for a fixed value of antiferromagnetic coupling, the transition temperature $T_c/K_B J_{\sigma-\sigma}$ increases. But when the antiferromagnetic coupling increases, for a fixed value of the ferromagnetic coupling, the transition temperature $T_c/K_B J_{\sigma-\sigma}$ decreases.

**Acknowledgements**

The authors gratefully acknowledge the generous financial support of CNRST Priority Program PPR 15/2015 and the European Union's Horizon 2020 research and innovation program under the Marie Skłodowska-Curie grant agreement No 778072.